\documentstyle{article}
\begin{document}
\addtolength{\topmargin}{-3.0cm}
\hsize=16cm
\textheight=23cm
\newtheorem{The}{Theorem}
\newcommand{\cc}{{\bf C}\ }
\newcommand{\beq}{\begin{equation} \label}
\newcommand{\eeq}{\end{equation}}
\newcommand{\pl}{\partial_}
\newcommand{\ve}{\varepsilon}
\newcommand{\vt}{\vartheta}
\newcommand{\vi}{\varphi}
\newcommand{\ol}{\overline}

\large
\begin{center}
{\LARGE Description  of  charged  scalar  system  by  means of General
Relativity approach}
\bigskip

{\Large Sergey    V.    Zuev}\footnote{Tel./Fax  +7.8412.441604,
E-mail: szuyev@tl.ru}\\
{\small \it Department  of  Theoretical   Physics,   Penza   State
Pedagogical University,    {\rm    440039}    Penza,   Russia}
\end{center}
\bigskip

\hrule
\medskip

\noindent {\bf \normalsize Abstract}
\medskip

\noindent {\normalsize  The K\"ahler metric which has been constructed
by present author in \cite{GP3} is used in this paper to find an exact
solution  of Einstein equations with energy-momentum tensor of special
type. That type  of  $T_{ij}$  admits  in  particular  to  use  it  as
energy-momentum tensor  of  charged  scalar  field and in case of such
system the field function is determined.}\\
\hrule
\bigskip

\bigskip\bigskip

{\bf \Large 1.} Let us consider a model consist of two particles where
one of them is the center of the inertial frame.  The problem to solve
is  to  determine  all  restrictions which appear by the gravitational
interaction reasoning.  These restrictions  must  be  applied  to  the
region of space-time in which the observable particle can move.

To start with we turn to the K\"ahler Ricci-flat 8-metric
\beq{metr}
g_{\alpha\ol{\beta}}=u''\ol{z^\alpha}z^\beta+u'\delta^\alpha_\beta=
{a^m\over     r^2}(r^m-a^m)^{\frac{1-m}{m}}\ol{z^\alpha}z^\beta+\frac{
(r^m-a^m)^{{1\over m}}}{r}\delta^\alpha_\beta,
\eeq
where
\beq{uprim}
u'\equiv{du\over dr}=\frac{(r^m-a^m)^{1/m}}{r},
\eeq
$u''\equiv du'/dr,$    $r\equiv    \sum\limits_{\alpha=1}^m   z^\alpha
\ol{z^\alpha},$ $m\equiv 4,$ $0<a={\rm const}.$
The metric (\ref{metr}) is determined in \cite{GP3} for  an  arbitrary
$m=2n.$ It  is  quite  simply  to  check  that for $m=4$ the metric is
hyperK\"ahler.

The first natural way to construct from  some  Riemannian  4-metric  a
Lorentz  4-metric  is so called "turn" transformation:  $x^4\to ix^4,$
where in case of 4-metric of type (\ref{metr}) $x^4={\rm Im} z^2.$ But
after  such turn the signature of $g$ becomes equal 0 or $\pm 4$ only.
We cannot find Lorentz metric in such a way.  By this reason we  start
from 8-metric  (\ref{metr})  which  has  notable  properties:  it  has
holomorphic isometry group equal to SU(4)  and  since  the  metric  is
K\"ahler and Ricci-flat its holonomy group is subgroup of SU(4).

Let us  come  back  to  the  model.  The  only  possible  way for time
measuring in the model is to use a frame clock (i.e.  clock which  has
been connected  with  frame  body).  Therefore  such  a  model must be
describe by means of space-time  model  which  become  flat  Minkovski
space in  case  of  gravity  absence.  The only invariant in Minkovski
space is so called interval
$$
r=c^2t^2-x^2-y^2-z^2.
$$
\bigskip

{\bf \Large 2.} "Turn" the gravity on now.  The question  is  how  the
geometric values of space-time depend on $r$? The interval $r$ must be
non-zero since the particle cannot  move  with  speed  of  light.  Let
$a\geq  0$  be  a  real  number  such  that  $|r|>a.$  Let  us  find a
pseudoRiemann metric in the connected region $U$ of  space-time  where
$|{x^4}^2-{x^3}^2-{x^2}^2-{x^1}^2|>a,$   $x^4=ct,$  $x^1=x,$  $x^2=y,$
$x^3=z.$

Using the  same  approach  that  in  classical two-body problem we can
formulate our problem in terms of central field  in  space-time.  This
means that all fields in the model must depend on $r$ only.  Moreover,
both matter and gravity fields must obey to  the  Einstein  equations.

Since the rest mass of elementary particle  is  little one  the  essential
gravity impact  may  be  just  in  case of ultra-relativistic particle
velosity, i.e.  $r$ is close to $a.$ In other words,  the metric  must
become Minkovski one when $r\to \infty$ (or $a\to 0$).

The isometry  group  of  Minkovski  metric  is Poincare group.  But in
space-time with central field  the  translations  are  prohibited  and
isometry group become Lorentz group.  If we suppose that region $U$ is
oriented that isometry group is SO(3,1).

For an arbitrary Lorentz  4-manifold  the  structure  group  of  frame
bundle is  O(3,1).  In  case  of region $U$ in this manifold has fixed
orientation the structure group on the $U$ is reduced to  SO(3,1).  In
the  same  time  it  is  well-known  that the holonomy group of simply
connected manifold is the subgroup of structure group of frame bundle.
That  is  why  the  restricted  holonomy  group  of $U$ is subgroup of
SO$(3,1).$
\bigskip

{\bf \Large  3.} Let us consider a complexification $U_c$ of $U,$ i.e.
the set which has the homeomorphysm to $U\times_\cc  U.$  The  complex
coordinates $(z^1,$ $\ldots,$ $z^4,$ $\ol{z^1},$ $\ldots,$ $\ol{z^4})$
on $U_c$ be  chosen  such  that  $z^j=x^j+iy^j,$  $\ol{z^j}=x^j-iy^j,$
$j=1,\ldots,4,$ where $x^j$ and $y^j$ are the real coordinates in $U.$
Then  we  can   construct   an   analytical   continuation   of   some
SO(3,1)-invariant  metric  $g$  on  $U$ to the Hermitian metric $h$ on
$U_c.$ The holomorphic isometries group of  $h$  is  SU(3,1)  and  its
restricted  holonomy  group  is subgroup of SU(3,1).  We can make some
kind  of  "turn"  transformation  now.  Let  us  put  $z^4=-z'^4$  and
$\ol{z^4}=\ol{z'^4}.$  Performing  such  transformation  everywhere we
find that the new Hermitian  metric  $h'$  has  SU(4)  as  holomorphic
isometries  group  and  its  restricted  holonomy group is subgroup of
SU(4).  We came to the conclusion that $h'$ has the same properties as
(\ref{metr}). By the Theorem 1 from \cite{GP3} the $h'$ is the same as
(\ref{metr}).

Making the inverse transformations we can easily find the  metric  $g$
on the  considered  region  $U$  of  space-time.  This  metric has the
following form.
\beq{5-3}
g_{ij}=u''x^kx^l\eta_k^i\eta_l^j+u'\eta^i_j,\quad i,j=1,\ldots,4,
\eeq
where
$$
\eta^i_j={\rm diag}(-1,-1,-1,1),
$$
\beq{5-r}
u'=\frac{(r^4-a^4)^{1/4}}{r},\quad u''=du'/dr,\quad
r=-(x^1)^2-(x^2)^2-(x^3)^2+(x^4)^2
\eeq
and have signature ($- - - +$) in case of $r>a$ or ($+++ -$) in case of
$r<-a.$

The following proposition is one of the result of the paper.
\begin{The}\label{T5-1}
Let $U$ be an oriented simply connected region in space-time and there
exist some coordinates $(x^1,x^2,x^3,x^4)$ in $U$ such that the metric
$g$  on  $U$ in these coordinates becomes Minkovski metric (up to sign
of signature) when $|{x^4}^2-{x^3}^2-{x^2}^2-{x^1}^2|\to \infty.$
Then $g$ has the {\rm  (\ref{5-3})} form.  Inverse, let on some region
$U$ of space-time the metric {\rm (\ref{5-3})}  exists.  Then  $U$  is
oriented and has a central field.
\end{The}
\bigskip

{\bf \Large 4.} The metric (\ref{5-3}) is not Ricci-flat.  That is way
the right side of Einstein equation is non-zero.  One can derive  that
$T_{ij}$ must be of the next form
\beq{5-5}
T_{ij}=P(r)g_{ij}+Q(r)\eta_{ij},
\eeq
where $P(r)$  and  $Q(r)$ are determined real fuctions on $r.$
\bigskip

Now let  us  consider for example proton and neutron as particles in
abovementioned model and let their spins have inverse directions. Then
total spin  of  the  system  is  zero and total electric charge of the
system is non-zero.  Therefore the system must be described by complex
scalar field.  The energy-momentum  tensor  of  complex  scalar  field
$\phi$ is
\beq{5-21}
T_{ik}=2\pl{i}\phi\pl{k}\ol{\phi}-g_{ik}m^2\phi\ol{\phi}
\eeq
and can     be     wrote    in    the    form    (\ref{5-5}),    where
$P(r)=8\phi'\ol{\phi'}+m^2\phi\ol{\phi},$ $Q(r)=-8\phi'\ol{\phi'}u'.$

Substitute (\ref{5-3}) and (\ref{5-21}) into Einstein equations
\beq{Ein}
G_{ij}\equiv r_{ij}-{1\over   2}\lambda   g_{ij}=\kappa   T_{ij},
\eeq
where $r_{ij}$ is Ricci tensor,  $\lambda$  is  scalar  curvature  and
$\kappa={8\pi G\over c^4}.$ This yields
$$
8\phi'\ol{\phi'}+m^2\phi\ol{\phi}u''=X(r),
$$
\beq{5-25}
\phi\ol{\phi}u'm^2=Y(r),
\eeq
where $\phi'={\pl\phi\over\pl r},$  $X$  and  $Y$  are  the  following
functions
$$
X={1\over 2}\frac{a^4(10r^8-7a^4r^4+3a^8)}{(r^4-a^4)^2r^6\kappa},
$$
$$
Y=-{1\over 2}\frac{a^4(r^4-3a^4)}{(r^4-a^4)r^5\kappa},
$$
Let
$$
{\rm Re}\,\phi\equiv \mu,\quad {\rm Im}\phi\equiv \nu.
$$
One can derive from (\ref{5-25}) that $\mu$  and  $\nu$  obey  to  the
following system of equations
$$
{\mu'}^2+{\nu'}^2={1\over 8}(X-Yu''/u')\equiv A(r),
$$
\beq{5-26}
\mu^2+\nu^2={Y\over m^2u'}\equiv B(r).
\eeq
The system has the next general solution.
$$
\mu=\pm F\cos(H+C),\quad \nu=\pm F\sin(H+C),
$$
where
$$
F= \sqrt{B},\quad H=\pm\int\sqrt{\frac{4AB-{B'}^2}{4B^2}}dr,
$$
$A$ and $B$ are  determined  by  (\ref{5-26}),  $C$  is  an  arbitrary
constant.

The $\phi$ has the following form
\beq{PHI}
\phi=F(r)e^{i[H(r)+C]}.
\eeq
The formulae  (\ref{5-3}) and (\ref{PHI}) expresses the exact solution
of Einstein equation for  two-body  charged  scalar  system  like  for
example neutron-proton or neutron-electron.

The geometrical  and  topological  structure  of  the  region  $U$  of
space-time and total space-time with considered system  is  not  fully
recognized now.  But  it  is clear that none particle can moves in the
region where $|r|\leq a.$

\end{document}